\documentclass[lettersize,journal]{IEEEtran}
\usepackage[flushleft]{threeparttable}
\usepackage{amsmath,amsfonts}
\usepackage{algorithmic}
\usepackage{algorithm}
\usepackage{array}
\usepackage[caption=false,font=normalsize,labelfont=sf,textfont=sf]{subfig}
\usepackage{textcomp}
\usepackage{stfloats}
\usepackage{url}
\usepackage{verbatim}
\usepackage{graphicx}
\usepackage{cite}
\begin{document}

\title{An \textbf{A}-$\Phi$ Formulation Solver in Electromagnetics\\ Based on Discrete Exterior Calculus}

\author{Boyuan Zhang, Dong-Yeop Na, Dan Jiao, and Weng Cho Chew
\\ School of Electrical and Computer Engineering\\Purdue University, West Lafayette, IN, 47907, US
\thanks{}
\thanks{}}



\maketitle

\begin{abstract}
An efficient numerical solver for the \textbf{A}-$\Phi$ formulation in electromagnetics based on discrete exterior calculus (DEC)
is proposed in this paper. The \textbf{A}-$\Phi$ formulation is immune to low-frequency breakdown and ideal for broadband and multi-scale analysis. The generalized Lorenz gauge is used in this paper, which decouples the \textbf{A} equation and the $\Phi$ equation. The \textbf{A}-$\Phi$ formulation is discretized by using the DEC, which is the discretized version of exterior calculus in differential geometry. In general, DEC can be viewed as a generalized version of the finite difference method, where Stokes' theorem and Gauss's theorem are naturally preserved. Furthermore, compared with finite difference method, where rectangular grids are applied, DEC can be implemented with unstructured mesh schemes, such as tetrahedral meshes. Thus, the proposed DEC \textbf{A}-$\Phi$ solver is inherently stable, free of spurious solutions and can capture highly complex structures efficiently. In this paper, the background knowledge about the \textbf{A}-$\Phi$ formulation and DEC is introduced, as well as technical details in implementing the DEC \textbf{A}-$\Phi$ solver with different boundary conditions. Numerical examples are provided for validation purposes as well.
\end{abstract}


\section{Introduction}
\IEEEPARstart{C}{omputational} electromagnetics (CEM) plays an important role in both academia and industry nowadays. There are two sets of formulations in CEM.
The first one is the \textbf{E}-\textbf{H} formulation in Maxwell's equations where the electric field \textbf{E} and magnetic field \textbf{H} are the unknowns to solve \cite{maxwell1865}.
The second is the \textbf{A}-$\Phi$ formulation, where \textbf{A} is the vector potential and $\Phi$ is the scalar potential of the electromagnetic field \cite{chew2014aphi}. The advantage of 
the \textbf{E}-\textbf{H} formulation is that its physical meaning is straightforward, since the electric field and magnetic field can be observed and measured quite easily.
However, due to the null space of the double curl operator in the \textbf{E}-\textbf{H} formulation, the discretized matrix equation of the \textbf{E}-\textbf{H} formulation
is badly conditioned, making it suffer from the low-frequency breakdown. Thus, when the frequency is low, or equivalently, when 
the object's size is much smaller than the wavelength, the \textbf{E}-\textbf{H} formulation must be solved with special care and with lower efficiency. In contrast, the
\textbf{A}-$\Phi$ formulation is immune to low-frequency breakdown thanks to the additional gauge term which removes the null space of the double curl operator. As a result,
the \textbf{A}-$\Phi$ formulation can be solved for both high frequencies and low frequencies with the same accuracy, making it ideal for broadband and multi-scale analysis \cite{Aphi2016inductance, yanlinlithesis, Yan2015Finite, 2018A}. More importantly, with the rapid development of quantum technology, the incorporation of quantum effects in computational electromagnetics is increasingly important. The vector potential \textbf{A} and scalar potential $\Phi$ are the natural bridges that connect classical electromagnetics and quantum mechanics, which is another advantage of the \textbf{A}-$\Phi$ formulation \cite{2020Quantum}.

Exterior calculus is a mathematical concept in differential geometry, which is similar to classical calculus, but in a more abstract sense. Differential geometry is proposed by E. Cartan in 1899 \cite{2004Differential}. In differential geometry, physical quantities are cast into differential forms, such as 0-forms, 1-forms and 2-forms. The inherent relations among differential forms are described by exterior calculus \cite{J2014On, 1991Differential}. The implementation of differential forms and exterior calculus in electromagnetics is pioneered by G. A. Deschamps in around 1981 \cite{Deschamps1981Electromagnetics}, where relevant concepts and their applications in electromagnetics are introduced. Discrete exterior calculus (DEC) is the discrete version of exterior calculus for computations. The term DEC was first introduced by A. N. Hirani in 2003 \cite{2003Discrete}, and some pioneering works of using the idea of DEC in CEM can be found in \cite{1998Computational, 2001DiscreteH, Tonti2002Finite, 1999Lattice}. Desbrun et. al. \cite{2005Discrete} further developed the concept of DEC in CEM. DEC can be thought as the generalized version of finite difference method (FDM), where both primal mesh and dual mesh are utilized \cite{2017Numerical, 2017ELECTROMAGNETIC}. There are many efforts in developing generalized finite difference method for non-rectilinear mesh and unstructured mesh, such as the cuvilinear FDM \cite{1992Finite}, contour and conformal FDM \cite{1997A}, discrete surface integral method (DSI) \cite{1995Divergence}, and generalized Yee algorithm \cite{1996Full}. Compared with the abovementioned generalized FDM, DEC is more concise, rigorous and easier to implement. In addtion, from the perspective of computation, DEC shares the same spirit with the finite integration technique (FIT), although the two methods originated from different communities \cite{2001discrete, 2004Recent}. In DEC, Stokes’ theorem, Gauss’s theorem and charge conservation are naturally preserved, which removes spurious solutions encountered in many numerical solvers \cite{2017Numerical, 2017ELECTROMAGNETIC}. Compared with FDM, which uses rectangular grids, flexible mesh schemes, such as unstructured mesh, can be applied in DEC, such as tetrahedral meshes. This enables DEC to capture complicated structures more easily and efficiently. Another widely used numerical solver in computational electromagnetics is finite element method (FEM)\cite{2015The}. Compared with FDM and DEC, FEM only uses primal mesh in its discretization. Thus, Stokes’ theorem and charge conservation cannot be guaranteed in FEM, which will sometimes give rise to spurious solutions, and special care must be paid to remove them \cite{Luis2011Spurious}. The incorporation of flexible mesh scheme and prime-dual grids in DEC makes it an efficient solver compared with other numerical algorithms.

In this paper, a DEC solver to the \textbf{A}-$\Phi$ formulation is proposed and its implementation details are provided as well. The rest of this paper is organized as follows.
In Section 2, the \textbf{A}-$\Phi$ formulation with generalized Lorenz gauge is introduced. In Section 3, details on applying DEC in solving the \textbf{A}-$\Phi$ formulation are provided. In Section 4, implementations of different boundary conditions in DEC are addressed. In Section 5, numerical examples are presented for validation purposes. Finally, in Section 6, the conclusion of this paper is summarized. In this paper, the time convention is $e^{-i \omega t}$.

\section{\textbf{A}-$\Phi$ Formulation}
The well-known Maxwell’s equations in frequency domain are
\begin{align}
\nabla \times \textbf{E}&=i \omega \textbf{B}, \label{deqn_ex1}\\
\nabla \times \textbf{H}&=-i \omega \textbf{D}+\textbf{J}, \label{deqn_ex2}\\
\nabla \cdot \textbf{D}&=\varrho, \label{deqn_ex3}\\
\nabla \cdot \textbf{B}&=0, \label{deqn_ex4}
\end{align}
where \textbf{E} and \textbf{H} are the electric field and magnetic field, respectively; \textbf{D} and \textbf{B} are electric flux density and magnetic flux density, respectively; 
\textbf{J} is the impressed current density; $\varrho$ is the impressed charge density; $\omega$ is the angular frequency of the time harmonic system, and $i$ is the imaginary unit.

The magnetic flux density \textbf{B} and electric field \textbf{E} can be expressed as

\begin{align}
\textbf{B} &= \nabla \times \textbf{A}, \label{deqn_ex5}\\
\textbf{E}&=i \omega \textbf{A}-\nabla \Phi. \label{deqn_ex8}
\end{align}

\textbf{A} and $\Phi$ above are the vector potential and scalar potential of the electromagnetic field. By plugging (\ref{deqn_ex5}) and (\ref{deqn_ex8}) into (\ref{deqn_ex2}) and (\ref{deqn_ex3}), respectively, and use the following constitutive relations in electromagnetics
\begin{align}
\textbf{D}&=\epsilon \textbf{E}, \label{deqn_ex9}\\
\textbf{B}&=\mu \textbf{H}, \label{deqn_ex10}
\end{align}
where $\epsilon$ and $\mu$ are the permittivity and permeability, respectively, one can obtain the equations for \textbf{A} and $\Phi$ as 
\begin{align}
\nabla \times \frac{1}{\mu} \nabla \times \textbf{A}-\omega^{2} \epsilon \textbf{A}-i\omega \epsilon \nabla \Phi &= \textbf{J}, \label{deqn_ex11}\\
\nabla \cdot (\epsilon \nabla \Phi)-\nabla \cdot (i \omega \epsilon \textbf{A}) &= -\varrho, \label{deqn_ex12}
\end{align}
In order to uniquely determine \textbf{A} and $\Phi$, proper gauge relation should be introduced to the above equations. In this paper, the generalized Lorenz gauge is used [1]. The generalized Lorenz gauge is valid for general media which can be inhomogeneous and lossy, and more importantly, the equations of \textbf{A} and $\Phi$ can be decoupled using the generalized Lorenz gauge. The generalized Lorenz gauge is \cite{chew2014aphi}:
\begin{align}
\nabla \cdot (\epsilon \textbf{A}) = i \omega \chi \Phi, \label{deqn_ex13}
\end{align}
where $\chi = \alpha \mu \epsilon^2$, and $\alpha$ is an arbitary constant. By using (\ref{deqn_ex13}) in (\ref{deqn_ex11}) and (\ref{deqn_ex12}), we have the following decoupled equations for \textbf{A} and $\Phi$:
\begin{align}
\nabla \times \frac{1}{\mu} \nabla \times \textbf{A}-\omega^2 \epsilon \textbf{A}-\epsilon \nabla [\chi^{-1}\nabla \cdot (\epsilon \textbf{A})]&=\textbf{J}, \label{deqn_ex14}\\
\nabla \cdot (\epsilon \nabla \Phi)+\omega^2\chi \Phi &= -\varrho, \label{deqn_ex15}
\end{align}
It should be noted that in non-static cases, (\ref{deqn_ex15}) can be derived from (\ref{deqn_ex14}) directly by taking divergence on both sides of (\ref{deqn_ex14}) with the general Lorenz gauge (\ref{deqn_ex13}), and by noticing the charge continuity equation
\begin{align}
\nabla \cdot \textbf{J} = i \omega \varrho. \label{deqn_ex16}
\end{align}
Thus, in non-static cases, one can either solve (\ref{deqn_ex15}) in tandem with (\ref{deqn_ex14}) to obtain the $\Phi$ result, or using the generalized Lorenz gauge (\ref{deqn_ex13}) to get $\Phi$ once the \textbf{A} equation is solved,

\begin{align}
\Phi = \frac{\nabla \cdot (\epsilon \mathbf{A})}{i \omega \chi}.
\end{align}

\section{Discrete Exterior Calculus in Electromagnetics}
As mentioned in the introduction, physical quantities are cast into differential forms in the context of differential geometry. Maxwell’s equations in terms of differential forms in frequency domain are written as \cite{J2014On, 2017Numerical, 2017ELECTROMAGNETIC}:
\begin{align}
dE &= i\omega B, \label{deqn_ex17}\\
dH &= -i\omega D+J, \label{deqn_ex18}\\
dD &=\varrho, \label{deqn_ex19}\\
dB &= 0, \label{deqn_ex20}
\end{align}
where $E$ and $H$ are 1-forms representing the electric field and magnetic field, respectively; $D$ and $B$ are 2-forms representing the electric flux density and magnetic flux density, respectively; the current density $J$ is a 2-form and the charge density $\varrho$ is a 3-form. In addition, the scalar potential $\Phi$ is a 0-form while the vector potential $A$ is a 1-form. The $d$ operator is the exterior derivative operator, which maps a $k$-form to a $(k+1)$-form.

Intuitively speaking, $k$-forms are quantities living on $k$-dimensional manifolds, and examples of $k$-dimensional manifolds are points, curves, surfaces, and volumes, when $k=0,1,2,3$.
When acting on 0, 1, 2-forms, the $d$ operator can be understood as gradient, curl and divergence operators, respectively. Since $k$-forms live on $k$-dimensional manifolds, they can be integrated using $k$-dimensional integrals.

DEC is the discretized version of exterior calculus, which makes computations with differential geometry possible. The idea of DEC originates from Whitney \cite{0Geometric}, who established the mathematical isomorphism between simplicial cochains and differential forms. An outstanding review paper on DEC can be found in \cite{2005Discrete}, where the history and detailed explanation of DEC are presented. 

\begin{figure}[!t]
\centering
\includegraphics[width=2.5in]{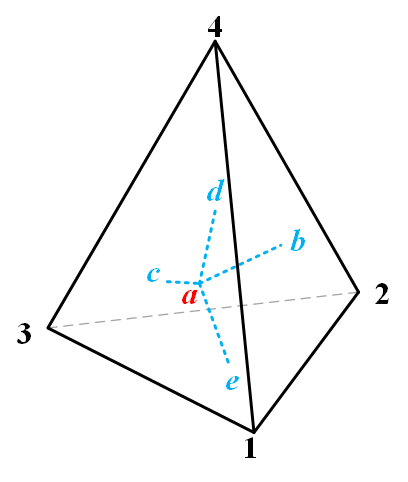}
\caption{An illustration of the primal tetrahedron and the dual mesh associated with it.}
\label{fig_1}
\end{figure}

We assume that a simplicial mesh is generated within the computation domain, such as tetrahedral meshes. Given the tetrahedral mesh, which is also referred to as the primal mesh, a dual mesh can be generated automatically by connecting the circumcenters or centroids of each tetrahedron and triangle in the primal mesh. In Fig. \ref{fig_1}, point $a$ is the circumcenter of tetrahedron $T_{1234}$ with vertices 1, 2, 3 and 4; points $b,c,d$ and $e$ are the circumcenters of triangles $S_{123},S_{134},S_{234}$ and $S_{124}$, respectively. Here, triangle $S_{ijk}$ means the vertices of the triangle are points $i$, $j$ and $k$ in Fig. \ref{fig_1}, where $i,j,k \in \{1,2,3,4\}$. By connecting the dual nodes (circumcenters of primal mesh), dual edges, dual surfaces and dual volumes can be constructed. Since the dual mesh is generated from the circumcenters of the primal mesh, in this paper, we denote such dual mesh as circumcenter-based dual mesh. Alternatively, we can also let points $a,b,c,d,e$ in Fig. \ref{fig_1} to be the centroids of the tetrahedron $T_{1234}$ and triangles $S_{123},S_{134},S_{234}$ and $S_{124}$, respectively. In this case, the dual nodes are determined using the centroids of the primal mesh, and thus such dual mesh is called centroid-based dual mesh. Note that if the dual mesh is generated based on circumcenters, each dual edge is orthogonal to the primal face associated with it, and vice versa. This is not true for centroid-based dual meshes. However, in 3D cases, circumcenters cannot be guaranteed to be well located inside their associated tetrahedrons/triangles, which will pose problems to the DEC solver. Thus, in 3D cases, centroid-based dual meshes are preferred. More detailed discussion about the two dual mesh strategies can be found in the Hodge star operator part in this section.

In electromagnetics, \textbf{E}, \textbf{B}, \textbf{A} and $\Phi$ are usually defined as primal forms, while \textbf{H}, \textbf{D}, \textbf{J} and $\varrho$ are usually defined as dual forms. Since \textbf{E} is a primal 1-form, in DEC, the line integral of \textbf{E} along each primal edge is defined as the unknown, which is called the cochain representation of differential forms \cite{J2014On, 2017Numerical, 2017ELECTROMAGNETIC}. For example, 
\begin{align}
E_i = \int_{l_i}\textbf{E} \cdot d\textbf{l}, \label{deqn_ex21}
\end{align}
is the integral of electric field along $i$-th primal edge $l_i$. Thus, the following vector, which includes the integral of electric field along each edge in the primal mesh, can be defined:
\begin{align}
\boldsymbol{E}=[E_1,E_2,\cdots,E_{N_1}]^T, \label{deqn_ex22}
\end{align}
where $N_1$ is the total number of edges in the primal mesh. Similarly, one can define
\begin{align}
A_i&=\int_{l_i}\textbf{A} \cdot d\textbf{l}, \label{deqn_ex23}\\
B_j&=\int_{S_j}\textbf{B} \cdot d\textbf{S}, \label{deqn_ex24}\\
\boldsymbol{\Phi}&=[\Phi_1,\Phi_2,\cdots,\Phi_{N_0}]^T, \label{deqn_ex25}\\
\boldsymbol{A}&=[A_1,A_2,\cdots,A_{N_1}]^T, \label{deqn_ex26}\\
\boldsymbol{B}&=[B_1,B_2,\cdots,B_{N_2}]^T, \label{deqn_ex27}
\end{align}
where $A_i$ is the line integral of \textbf{A} along $i$-th primal edge $l_i$; $\Phi_n$ is the scalar potential on $n$-th primal node $p_n$ (integration of 0-forms); $N_0$ is the total number of primal nodes; $S_j$ is the $j$-th primal surface in the mesh; $B_j$ is the magnetic flux on $S_j$; $N_2$ is the total number of primal surfaces. Vector $\boldsymbol{\Phi}$ is called a primal 0-cochain since it is the discretized version of 0-forms, $\boldsymbol{E}$ and $\boldsymbol{A}$ are called primal 1-cochains, and $\boldsymbol{B}$ is called primal 2-cochain.

For the dual form quantities  \textbf{H}, \textbf{D}, \textbf{J} and $\varrho$, similar definitions can be made on dual edges, dual surfaces, and dual volumes. Specifically, for dual 3-form $\varrho$, its dual 3-cochain representation is
\begin{align}
\varrho_i&=\int_{V_i^{\text{dual}}}\varrho \cdot dV, \label{deqn_ex28}\\
\boldsymbol{\varrho}&=[\varrho_1,\varrho_2,\cdots,\varrho_{N_0}]^T, \label{deqn_ex29}
\end{align}
where $V_i^{\text{dual}}$ is the $i$-th dual volume (which is associated with the $i$-th primal node $p_i$), and $N_0$ is the total number of dual volume (equals to the total number of primal node).

The $d$ operator is represented by the incidence matrices in DEC, namely, $\overline{\mathbf{d}}^{(0)},\overline{\mathbf{d}}^{(1)}$ and $\overline{\mathbf{d}}^{(2)}$. The dimension of  $\overline{\mathbf{d}}^{(0)}$ is $N_1 \times N_0$, and the superscript $(0)$ indicates $\overline{\mathbf{d}}^{(0)}$ operates on primal 0-cochains. The $(i, j)$ element of  $\overline{\mathbf{d}}^{(0)}$ is
\begin{align}
\left[\overline{\mathbf{d}}^{(0)}\right]_{i,j}=d_{i,j}^{(0)}= \begin{cases}
\pm 1, 
\text{ if $p_j$ is a vertex of $l_i$}\\
0,  \text{ otherwise }
\end{cases}
, \label{deqn_ex30}
\end{align}
where $p_j$ is the $j$-th primal node and $l_i$ is the $i$-th primal edge. 
Similarly, $\overline{\mathbf{d}}^{(1)}$ is an $N_2 \times N_1$ matrix, and $\overline{\mathbf{d}}^{(2)}$ is an $N_3 \times N_2$ matrix, where $N_3$ is the total number of primal volumes. Their entries are:
\begin{align}
\left[\overline{\mathbf{d}}^{(1)}\right]_{i,j}=d_{i,j}^{(1)}= \begin{cases}
\pm 1, 
\text{ if $l_j$ is an edge of $S_i$}\\
0,  \text{ otherwise }
\end{cases}
, \label{deqn_ex31}
\end{align}
\begin{align}
\left[\overline{\mathbf{d}}^{(2)}\right]_{i,j}=d_{i,j}^{(2)}= \begin{cases}
\pm 1, 
\text{ if $S_j$ is an edge of $V_i$}\\
0,  \text{ otherwise }
\end{cases}
, \label{deqn_ex32}
\end{align}
where $S_i$ is the $i$-th primal surface; $V_i$ is the $i$-th primal volume. The sign of the $\pm 1$ entries in (\ref{deqn_ex30})-(\ref{deqn_ex32}) is determined by orientation. In DEC, edges, surfaces and volumes are with positive orientations. If a node $p_j$ is the start/end point of edge $l_i$, then $d_{i,j}^{(0)}=1/-1$; if edge $l_j$ and surface $S_i$ have the same/opposite orientation (see \cite{2017ELECTROMAGNETIC}), $d_{i,j}^{(1)}=1/-1$; if the positive direction of surface $S_j$ points outside/inside to volume $V_i$, $d_{i,j}^{(2)}=1/-1$. In general, $\overline{\mathbf{d}}^{(k)}$ matrix acts on a primal $k$-cochain, and gives back a primal $(k+1)$-cochain.

It can be shown that $\overline{\mathbf{d}}^{(0)}$, $\overline{\mathbf{d}}^{(1)}$ and $\overline{\mathbf{d}}^{(2)}$ represent gradient, curl and divergence operators in the discrete mesh. For example, the integral form of (\ref{deqn_ex4}) on a tetrahedron is
\begin{align}
\oint_{\partial V_i}\textbf{B} \cdot d \textbf{S} = \sum_{S_j \in \partial V_i}\int_{S_j} \textbf{B} \cdot d \textbf{S} = \sum_{S_j \in \partial V_i} d_{i,j}^{(2)} B_j=0,
\label{deqn_ex33} 
\end{align}
where $\partial V_i$ is the boundary of the primal volume $V_i$; $B_j$ is the $j$-th element in vector $\boldsymbol{B}$. Thus, the matrix equation of (\ref{deqn_ex4}) in DEC is
\begin{align}
\overline{\mathbf{d}}^{(2)}\boldsymbol{B}=0. \label{deqn_ex34}
\end{align}
Similarly, (\ref{deqn_ex1})-(\ref{deqn_ex3}) can be expressed as DEC matrix equations, and the Maxwell’s equations in DEC form are \cite{2017Numerical, 2017ELECTROMAGNETIC}:
\begin{align}
\overline{\mathbf{d}}^{(1)}\boldsymbol{E}&=i\omega \boldsymbol{B},\label{deqn_ex35}\\
\overline{\mathbf{d}}^{(1)}_{\text{dual}}\boldsymbol{H}&=\left[\overline{\mathbf{d}}^{(1)}\right]^T\boldsymbol{H}=-i\omega \boldsymbol{D}+\boldsymbol{J},\label{deqn_ex36}\\
\overline{\mathbf{d}}^{(2)}_{\text{dual}}\boldsymbol{D}&=-\left[\overline{\mathbf{d}}^{(0)}\right]^T\boldsymbol{D}=\boldsymbol{\varrho},\label{deqn_ex37}\\
\overline{\mathbf{d}}^{(2)}\boldsymbol{B}&=0, \label{deqn_ex38}
\end{align}
where $\overline{\mathbf{d}}_{\text{dual}}^{(\cdot)}$ are the incidence matrices on the dual mesh. For example, $\overline{\mathbf{d}}_{\text{dual}}^{(0)}$ contains the connectivity information among dual edges and dual nodes

\begin{align}
\left[\overline{\mathbf{d}}_{\text{dual}}^{(0)}\right]_{i,j}= \begin{cases}
\pm 1, 
\text{ if $p^{\text{dual}}_j$ is a vertex of $l^{\text{dual}}_i$}\\
0,  \text{ otherwise }
\end{cases}
, \label{deqn_ex38.5}
\end{align}
where $p^{\text{dual}}_j$ is the dual node associated with primal volume $V_j$, $l^{\text{dual}}_i$ is the dual edge associated with primal surface $S_i$, the $\pm$ sign is determined by orientation. The other two dual incidence matrices $\overline{\mathbf{d}}_{\text{dual}}^{(1)}$ and $\overline{\mathbf{d}}_{\text{dual}}^{(2)}$ can be interpreted similarly. Note that the dual nodes, edges, surfaces and volumes are indexed according to the indices of their associated primal volumes, surfaces, edges and nodes, respectively. In addition, the dual edges and surfaces share the same orientations with the relevant primal surfaces and edges \cite{1999Lattice}. Due to the complementary nature between the primal mesh and dual mesh, we have 

\begin{align}
\overline{\mathbf{d}}_{\text{dual}}^{(n-k)}=(-1)^k\left[\overline{\mathbf{d}}^{(k-1)}\right]^T, \label{deqn_ex39}
\end{align}
where $n$ is the dimension of the problem. In 3D cases, $n=3$, $k=1,2,3$.

The constitutive relations (\ref{deqn_ex9}) and (\ref{deqn_ex10}), and the generalized Lorenz gauge (\ref{deqn_ex13}) can be represented by the following matrix equations in DEC:
\begin{align}
\boldsymbol{D}&=\mathbf{\star}_{\epsilon}^{(1)}\boldsymbol{E}, \label{deqn_ex40}\\
\boldsymbol{H}&=\mathbf{\star}_{\mu ^{-1}}^{(2)}\boldsymbol{B}, \label{deqn_ex41}\\
\overline{\mathbf{d}}_{\text{dual}}^{(2)}\left(\mathbf{\star}_{\epsilon}^{(1)}\boldsymbol{A}\right)&=-\left(\overline{\mathbf{d}}^{(0)}\right)^T\left(\mathbf{\star}_{\epsilon}^{(1)}\boldsymbol{A}\right)=i\omega\mathbf{\star}_{\chi}^{(0)}\boldsymbol{\Phi}, \label{deqn_ex42}
\end{align}
where $\mathbf{\star}_{\epsilon}^{(1)}$, $\mathbf{\star}_{\mu^{-1}}^{(2)}$ and $\mathbf{\star}_{\chi}^{(0)}$ are called Hodge star operators, and the superscript $(k)$ means they act on primal/dual $k$-cochains. The Hodge star operators describe the relations between primal forms/cochains and dual forms/cochains. In the above equations, $\mathbf{\star}_{\epsilon}^{(1)}$ maps primal 1-cochain $\boldsymbol{E}$ to dual 2-cochain $\boldsymbol{D}$; $\mathbf{\star}_{\mu^{-1}}^{(2)}$ maps primal 2-cochain $\boldsymbol{B}$ to dual 1-cochain $\boldsymbol{H}$; $\mathbf{\star}_{\chi}^{(0)}$ maps primal 0-cochain $\boldsymbol{\Phi}$ to dual 3-cochain $\overline{\mathbf{d}}_{\text{dual}}^{(2)}\left(\mathbf{\star}_{\epsilon}^{(1)} \boldsymbol{A}\right)$.

In general, there are two ways to construct the Hodge star operators in DEC. The first is based on circumcenter dual mesh, as described in \cite{J2014On, 2017Numerical}. Since the circumcenter-based dual mesh is completely orthogonal to the primal mesh, the Hodge star operators are diagonal matrices with very straightforward interpretation. However, it is very difficult to guarantee all circumcenters are well located in their tetrahedrons/triangles in 3D cases, which poses great difficulty in using the circumcenter-based formulation. To tackle this problem, \cite{HOT} proposed a Hodge-optimized triangulation meshing scheme, which utilizes a weighted-circumcenter concept to guarantee all the circumcenters are well located. The second way is to use non-diagonal Hodge star operators. In this paper, the non-diagonal Galerkin Hodge star operator is used which utilizes the centroid-based dual mesh and Whitney forms \cite{2015The, 1999Some, 2006Geometric}. The Galerkin Hodge star operators can be expressed as follows:
\begin{align}
\left[\mathbf{\star}_{\epsilon}^{(1)}\right]_{i,j}&=\left<\textbf{W}_i^1,\epsilon \cdot \textbf{W}_j^1 \right>, \label{deqn_ex43}\\
\left[\mathbf{\star}_{\mu^{-1}}^{(2)}\right]_{i,j}&=\left<\textbf{W}_i^2,\mu^{-1} \cdot \textbf{W}_j^2 \right>, \label{deqn_ex44}\\
\left[\mathbf{\star}_{\chi}^{(0)}\right]_{i,j}&=\left<W_i^0,\chi \cdot W_j^0 \right>, \label{deqn_ex45}
\end{align}
where $\textbf{W}_i^1$ is the Whitney 1-form associated with primal edge $l_i$; $\textbf{W}_i^2$ is the Whitney 2-form associated with primal surface $S_i$; $W_i^0$ is the Whitney 0-form associated with primal node $p_i$. $\epsilon$, $\mu$ and $\chi$ are piecewise constant parameters within each tetrahedron. Specifically, suppose a tetrahedron in the mesh is the one in Fig. \ref{fig_1}, $T_{1234}$. For $T_{1234}$, the Whitney 0-form associated with node $m$, where $m=1,2,3,4$, is
\begin{align}
W_m^0(\mathbf{r})=\lambda_m(\mathbf{r}),
\end{align}
where $\lambda_m(\mathbf{r})$ is the barycentric coordinate of point $\mathbf{r}$ with respect to node $m$, and point $\mathbf{r}$ is inside $T_{1234}$.

The Whitney 1-form associated with edge $l_{mn}$, where the orientation of $l_{mn}$ is from node $m$ to node $n$, and $m,n \in \{1,2,3,4\}$, $m \neq n$, is
\begin{align}
\textbf{W}_{mn}^1(\mathbf{r}) = \lambda_m(\mathbf{r}) \nabla \lambda_n(\mathbf{r}) - \lambda_n(\mathbf{r}) \nabla \lambda_m(\mathbf{r}),
\end{align}
where $\lambda_m(\mathbf{r})$ and $\lambda_n(\mathbf{r})$ are the Whitney 0-forms associated with node $m$ and $n$ in $T_{1234}$, and point $\mathbf{r}$ is inside $T_{1234}$.

The Whitney 2-form associated with surface $S_{mnp}$, where $m,n,p \in \{1,2,3,4\}$, $m,n,p$ are distinct, and the orientation of $S_{mnp}$ is determined by the order $m \to n \to p$ using the right-hand rule, is
\begin{multline}
\mathbf{W}_{mnp}^2(\mathbf{r})= 2(\lambda_m(\mathbf{r}) \nabla \lambda_n(\mathbf{r}) \times \nabla \lambda_p(\mathbf{r})+ \\
\lambda_n(\mathbf{r}) \nabla \lambda_p(\mathbf{r}) \times \nabla \lambda_m(\mathbf{r}) +\lambda_p(\mathbf{r}) \nabla \lambda_m(\mathbf{r}) \times \nabla\lambda_n(\mathbf{r})).
\end{multline}

Thus, Whitney forms can be considered as local basis functions within each tetrahedron in the mesh, which can be determined by the coordinates of the four nodes of the tetrahedron and the orientation of the mesh.

Similar to the DEC Maxwell’s equations, the $\textbf{A}$-$\Phi$ equations can be discretized using DEC as well:
\begin{multline}
\left(\overline{\mathbf{d}}^{(1)}\right)^T \mathbf{\star}_{\mu^{-1}}^{(2)} \overline{\mathbf{d}}^{(1)} \boldsymbol{A} - \omega^2 \mathbf{\star}_{\epsilon}^{(1)} \boldsymbol{A} +\\
\mathbf{\star}_{\epsilon}^{(1)} \overline{\mathbf{d}}^{(0)} \mathbf{\star}_{\chi^{-1}}^{(3)} \left( \overline{\mathbf{d}}^{(0)}\right)^T \mathbf{\star}_{\epsilon}^{(1)} \boldsymbol{A}
=\boldsymbol{J}, \label{deqn_ex46}
\end{multline}
\begin{align}
-\left(\overline{\mathbf{d}}^{(0)}\right)^T \mathbf{\star}_{\epsilon}^{(1)} \overline{\mathbf{d}}^{(0)} \boldsymbol{\Phi}+\omega^2 \mathbf{\star}_{\chi}^{(0)} \boldsymbol{\Phi} = -\boldsymbol{\varrho}.
\label{deqn_ex47}
\end{align}
Note that $\mathbf{\star}_{\chi^{-1}}^{(3)}$ is the inverse of $\mathbf{\star}_{\chi}^{(0)}$. Since $\mathbf{\star}_{\chi}^{(0)}$ is sparse, its exact inverse is in general a dense matrix, which is not favored in solving large scale problems. To maintain the sparsity of (\ref{deqn_ex46}), one can use the sparse approximate inverse (SPAI) technique \cite{Grote1997Parallel} to compute the sparse approximate of $\mathbf{\star}_{\chi^{-1}}^{(3)}$. Alternatively, one can also approximately construct $\mathbf{\star}_{\chi}^{(0)}$ using its geometric interpretation as:
\begin{align}
\left[\mathbf{\star}_{\chi}^{(0)}\right]_{i,j}= \begin{cases}
\chi V_i^{\text{dual}}, 
\text{ if $i=j$}\\
0,  \text{ otherwise }
\end{cases}
, \label{deqn_ex48}
\end{align}
where $V_i^{\text{dual}}$ is the volume of the $i$-th dual mesh. It is easy to see that this approximate $\mathbf{\star}_{\chi}^{(0)}$ is diagonal, and thus, its inverse is trivial to find:
\begin{align}
\left[\mathbf{\star}_{\chi^{-1}}^{(3)}\right]_{i,j}= \begin{cases}
\frac{1}{\chi V_i^{\text{dual}}}, 
\text{ if $i=j$}\\
0,  \text{ otherwise }
\end{cases}
. \label{deqn_ex49}
\end{align}
Now both (\ref{deqn_ex46}) and (\ref{deqn_ex47}) are sparse matrix equations. In addition, from the charge continuity equation (\ref{deqn_ex16}), we have:
\begin{align}
-\left(\overline{\mathbf{d}}^{(0)}\right)^T \boldsymbol{J} = i \omega \boldsymbol{\varrho}. \label{deqn_ex50}
\end{align}

When solving the DEC $\textbf{A}$-$\Phi$ equations (\ref{deqn_ex46}) and (\ref{deqn_ex47}), one can use (\ref{deqn_ex50}) to solve the two equations in tandem. After the vector potential $\boldsymbol{A}$ and scalar potential $\boldsymbol{\Phi}$ cochains are solved, the electric field $\boldsymbol{E}$ and magnetic flux density $\boldsymbol{B}$ cochains can be extracted from $\boldsymbol{A}$ and $\boldsymbol{\Phi}$ by using (\ref{deqn_ex8}) and (\ref{deqn_ex5}). Specifically, in DEC, $\boldsymbol{E}$ and $\boldsymbol{B}$ can be obtained by:
\begin{align}
\boldsymbol{E}&=i \omega \boldsymbol{A} - \overline{\mathbf{d}}^{(0)} \boldsymbol{\Phi}, \label{deqn_ex51}\\
\boldsymbol{B}&= \overline{\mathbf{d}}^{(1)} \boldsymbol{A}.  \label{deqn_ex52}
\end{align}

\section{Boundary Conditions}

Due to the numerical truncation of the calculation domain and the incomplete dual mesh at the truncated boundary, proper boundary conditions need to be implemented in DEC. In this section, the implementation of perfect magnetic conductor (PMC), perfect electric conductor (PEC), periodic boundary condition (PBC) and absorbing boundary condition such as perfectly matched layers (PML) are illustrated.

\subsection{Default Boundary Condition: PMC}
\begin{figure}[!t]
\centering
\includegraphics[width=2.5in]{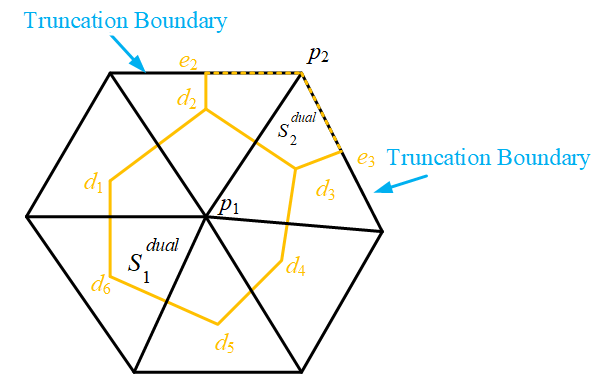}
\caption{Illustration of the incompleteness of the dual mesh in 2D case due to the numerical truncation.}
\label{fig_2}
\end{figure}

The incompleteness of the dual mesh due to the boundary truncation is buried in the incidence matrices of the dual mesh \cite{2017Numerical}, namely, $\overline{\mathbf{d}}_{\text{dual}}^{(k)}$. Fig. \ref{fig_2} shows an example of the incomplete dual mesh in 2D case for a better visualization. In (\ref{deqn_ex46}), the first term can be written as 
\begin{align}
\left(\overline{\mathbf{d}}^{(1)}\right)^T \mathbf{\star}_{\mu^{-1}}^{(2)} \overline{\mathbf{d}}^{(1)} \boldsymbol{A}=\left(\overline{\mathbf{d}}^{(1)}\right)^T \boldsymbol{H}. \label{deqn_ex53}
\end{align}
Since $\boldsymbol{H}$ is a dual 1-cochain and $\left(\overline{\mathbf{d}}^{(1)}\right)^T \boldsymbol{H}$ gives a dual 2-cochain on dual surface, (\ref{deqn_ex53}) is equivalent to summing up $H_i$ on the dual edges of each dual surface. Here, $H_i$ denotes the $i$-th element in the $\boldsymbol{H}$ cochain, namely, the integral of magnetic field $\mathbf{H}$ along the $i$-th dual edge. For example, the $j$-th element in the dual 2-cochain vector $\left(\overline{\mathbf{d}}^{(1)}\right)^T \boldsymbol{H}$ can be expressed as:
\begin{align}
\left[ \left( \overline{\mathbf{d}}^{(1)}\right)^T \boldsymbol{H} \right]_j = \sum_{l_i^{\text{dual}} \in \partial S_j^{\text{dual}}} H_i. \label{deqn_ex54}
\end{align}
where $\partial S_j^{\text{dual}}$ represents the dual edges forming the dual surface $S_j^{\text{dual}}$. If the above expression is performed with respect to $S_1^{\text{dual}}$ and $S_2^{\text{dual}}$ in Fig. \ref{fig_2}, where $S_1^{\text{dual}}$ is composed of nodes $d_1$, $d_2$, $d_3$, $d_4$, $d_5$ and $d_6$; $S_2^{\text{dual}}$ is composed of nodes $d_2$, $e_2$, $p_2$, $e_3$ and $d_3$, we have
\begin{multline}
\left[ \left( \overline{\mathbf{d}}^{(1)} \right)^T \boldsymbol{H} \right]_1 = H_{d_1 d_2}+H_{d_2 d_3}+H_{d_3 d_4}\\+H_{d_4 d_5}+H_{d_5 d_6}+H_{d_6 d_1}, \label{deqn_ex55}
\end{multline}
\begin{align}
\left[ \left( \overline{\mathbf{d}}^{(1)} \right)^T \boldsymbol{H} \right]_2 = H_{d_2 d_3}+H_{d_2 e_2}+H_{d_3 e_3}. \label{deqn_ex56}
\end{align}
where $H_{d_1 d_2}$ is the integral of the magnetic field $\mathbf{H}$ along the dual edge $l_{d_1 d_2}^{\text{dual}}$, and other terms can be interpreted similarly. As can be seen, since the boundary dual edges $l_{e_2 p_2}^{\text{dual}}$ and $l_{p_2 e_3}^{\text{dual}}$ are not included in the dual incidence matrix $\left( \overline{\mathbf{d}}^{(1)} \right)^T$, it is implied that $H_{e_2 p_2}$ and $H_{p_2 e_3}$ are equal to zero, which is the PMC boundary condition. Thus, the PMC boundary condition is embedded in the DEC matrix equation. Without other boundary condition imposed, PMC is the default boundary condition in DEC.

\subsection{PEC Boundary Condition}

The boundary condition for PEC in $\textbf{E}$-$\textbf{H}$ formulation is the tangential component of the electric field, $E_t$, equals zero. From (\ref{deqn_ex8}), it can be seen that for $\textbf{A}$-$\Phi$ formulation, the PEC boundary condition requires the tangential component of the vector potential, $A_t$, is zero, and the scalar potential $\Phi$ is a constant on the PEC surfaces. In DEC, $\textbf{A}$ cochains on PEC boundary primal edges must be zero, and primal node $\Phi$ cochains on each PEC body are the same constant. In addition, since the $\textbf{A}$ equation (\ref{deqn_ex14}) is actually the Ampere’s law (\ref{deqn_ex2}) on each dual surface and the $\Phi$ equation (\ref{deqn_ex15}) is the Gauss’s law (\ref{deqn_ex3})  on each dual volume of the electric flux density $\textbf{D}$, primal boundary nodes, edges and surfaces should be excluded from the DEC matrix equation, i.e., from the incidence matrices $\overline{\mathbf{d}}^{(0)}$, $\overline{\mathbf{d}}^{(1)}$ and $\overline{\mathbf{d}}^{(2)}$. Otherwise, if the primal boundary nodes, edges and surface are included, it implies that the tangential magnetic field $H_t$ and the electric flux density $\textbf{D}$ on the PEC surface is zero, which is not true due to the surface current and charge on the PEC boundary. The reason here is similar to that outlined in the PMC boundary condition: the dual mesh is incomplete due to the truncation of the PEC boundary. To accommodate PEC boundary condition in DEC, one can perform the following manipulations on the incidence matrices:
\begin{enumerate}
\item{Find the indices of the primal boundary nodes, edges and surface on the PEC boundary, which are $I_n$, $I_e$ and $I_s$, repectively.}
\item{Remove columns in $\overline{\mathbf{d}}^{(0)}$ whose indices belong to $I_n$. Those columns represent primal boundary nodes.}
\item{Remove rows in $\overline{\mathbf{d}}^{(0)}$ whose indices belong to $I_e$. Those rows represent primal boundary edges.}
\item{Remove columns in $\overline{\mathbf{d}}^{(1)}$ whose indices belong to $I_e$. Those columns represent primal boundary edges.}
\item{Remove rows in $\overline{\mathbf{d}}^{(1)}$ whose indices belong to $I_s$. Those rows represent primal boundary surfaces.}
\item{Remove columns in $\overline{\mathbf{d}}^{(2)}$ whose indices belong to $I_s$. Those columns represent primal boundary surfaces.}
\end{enumerate}

After the above manipulations, the dimension of $\overline{\mathbf{d}}^{(0)}$, $\overline{\mathbf{d}}^{(1)}$ and $\overline{\mathbf{d}}^{(2)}$ are $N_{1,\text{in}} \times N_{0,\text{in}}$, $N_{2,\text{in}} \times N_{1,\text{in}}$ and $N_3 \times N_{2,\text{in}}$, respectively, where $N_{0,\text{in}}$, $N_{1,\text{in}}$ and $N_{2,\text{in}}$ are the number of primal nodes, edges and surfaces strictly inside the PEC boundary. The new $\overline{\mathbf{d}}^{(0)}$,  $\overline{\mathbf{d}}^{(1)}$ and  $\overline{\mathbf{d}}^{(2)}$ represent the connectivity among these inner primal nodes, edges and surfaces. When constructing the dual incidence matrices in (\ref{deqn_ex39}), the new $\overline{\mathbf{d}}^{(0)}$,  $\overline{\mathbf{d}}^{(1)}$ and  $\overline{\mathbf{d}}^{(2)}$ should be used accordingly as well.

Since the dimensions of $\overline{\mathbf{d}}^{(0)}$,  $\overline{\mathbf{d}}^{(1)}$ and  $\overline{\mathbf{d}}^{(2)}$ have been reduced, the Hodge star matrices in (\ref{deqn_ex43})-(\ref{deqn_ex45}) should be updated as well to accommodate for the new incidence matrices. Only the inner primal nodes, edges and surfaces should be taken into account when constructing the Hodge star matrices. It should also be noticed that the indices of the primal nodes, edges and surfaces have been changed according to the row and column indices of the new incidence matrices $\overline{\mathbf{d}}^{(0)}$,  $\overline{\mathbf{d}}^{(1)}$ and  $\overline{\mathbf{d}}^{(2)}$, since the primal boundary rows/columns in the incidence matrices have been removed.

With the above manipulations, only the dual meshes that are strictly inside the PEC boundary with no truncation are considered, and the zero tangential electric field along the PEC surface is implied. 

\subsection{Periodic Boundary Condition}
\begin{figure}[!t]
\centering
\includegraphics[width=2.5in]{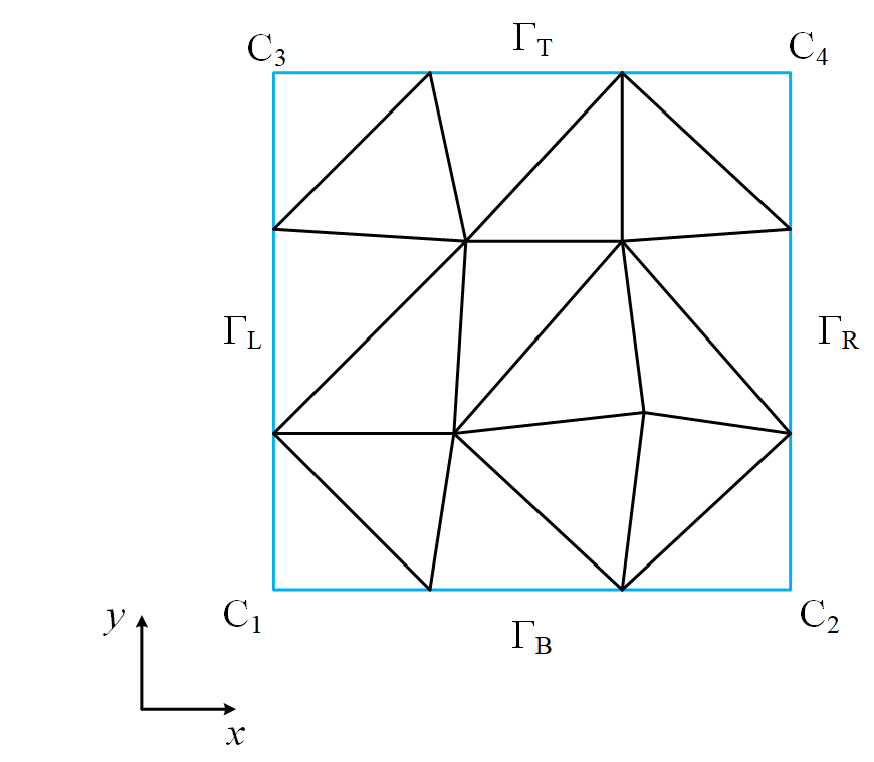}
\caption{Illustration on symmetric boundary mesh in 2D case.}
\label{fig_3}
\end{figure}

The periodic boundary condition (PBC) is widely used in physics, such as in the study of photonic crystals, and typically implemented in eigenvalue problems. The eigenmodes in a 2D photonic crystal can be characterized using the Bloch's theorem, where the eigenmode is a planewave modulated by a periodic function\cite{joannopoulos2008molding}. Fig. \ref{fig_3} shows an example of a unit cell of photonic crystal with symmetric mesh in 2D case. Suppose the length of the unit cell along the $x$ and $y$ directions are $L_x$ and $L_y$, respectively, the eigenmode of $\mathbf{A}$ can be written as:
\begin{align}
\mathbf{A} = e^{i\mathbf{k} \cdot \mathbf{r}} \mathbf{u}_{\mathbf{k}}(x,y),  \label{deqn_ex56.5}
\end{align}
where $\mathbf{k}$ is the wave vector of the plane wave, $\mathbf{k}=k_x \hat{x}+k_y \hat{y}$; $\mathbf{r}$ is the position vector in the 2D plane; $\mathbf{u}_{\mathbf{k}}(x,y)$ is a periodic function with respect to the primitive lattice vector $\mathbf{R}=(L_x, L_y)$, i.e., $\mathbf{u}_{\mathbf{k}}(x,y) = \mathbf{u}_{\mathbf{k}}(x+mL_x,y+nL_y)$, where $m$ and $n$ are arbitary integers.

In order to implement PBC in DEC, the mesh must be symmetric on the periodic boundaries.  In the 2D case, the vector potential cochain $\boldsymbol{A}$ can be divided into five blocks:
\begin{align}
\boldsymbol{A}=\left[\boldsymbol{A}_I,\boldsymbol{A}_{\Gamma_L},\boldsymbol{A}_{\Gamma_R},\boldsymbol{A}_{\Gamma_T},\boldsymbol{A}_{\Gamma_B} \right], \label{deqn_ex57}
\end{align}
where $\boldsymbol{A}_I$ denotes the $\textbf{A}$ cochains on the inner primal edges, $\boldsymbol{A}_{\Gamma_L},\boldsymbol{A}_{\Gamma_R},\boldsymbol{A}_{\Gamma_T},\boldsymbol{A}_{\Gamma_B}$ are the $\textbf{A}$ cochains on the four periodic boundaries in Fig. \ref{fig_3}. From the PBC and the periodic property of the Bloch mode (\ref{deqn_ex56.5}), we have
\begin{align}
\boldsymbol{A}_{\Gamma_R} = e^{-i \Psi_x} \boldsymbol{A}_{\Gamma_L}, \label{deqn_ex58}\\
\boldsymbol{A}_{\Gamma_T} = e^{-i \Psi_y} \boldsymbol{A}_{\Gamma_B}, \label{deqn_ex59}
\end{align}
where $\Psi_x = k_x L_x$ and $\Psi_y = k_y L_y$ are the phase shifts between adjacent unit cells along $x$ and $y$ directions; $L_x$ and $L_y$ are the length of the cell of interest in Fig. \ref{fig_3} along $x$ and $y$ directions; $k_x$ and $k_y$ are the $x$ component and $y$ component of the wave vector $\mathbf{k}$. We can define a reduced $\textbf{A}$ cochain and a projection matrix $\overline{\mathbf{P}}_A$ as \cite{2017Numerical}
\begin{align}
\boldsymbol{A}_{\text{reduced}} = \left[ \boldsymbol{A}_I, \boldsymbol{A}_{\Gamma_L},\boldsymbol{A}_{\Gamma_B} \right], \label{deqn_ex60}
\end{align}
\begin{align}
\overline{\mathbf{P}}_A = 
\begin{bmatrix}
\overline{\mathbf{I}} & 0 &0 & 0 \\
0 & \overline{\mathbf{I}} & 0 & 0 \\
0 & \overline{\mathbf{I}}e^{-i \Psi_x} & 0 & 0\\
0 & 0 & \overline{\mathbf{I}} & 0 \\
0 & 0 & \overline{\mathbf{I}}e^{-i \Psi_y} & 0\\
0 & 0 & 0 & \overline{\mathbf{I}}
\end{bmatrix}
, \label{deqn_ex61}
\end{align}
where $\overline{\mathbf{I}}$ is the identity matrix. Thus, we have
\begin{align}
\boldsymbol{A} = \overline{\mathbf{P}}_A \boldsymbol{A}_{\text{reduced}}. \label{deqn_ex62}
\end{align}
Thus, to accommodate the PBC, the generalized eigenvalue problem from (\ref{deqn_ex46}) and (\ref{deqn_ex47}) can be written as
\begin{multline}
\left( \overline{\mathbf{P}}_A \right)^H \left[ \left( \overline{\mathbf{d}}^{(1)} \right)^T \mathbf{\star}_{\mu^{-1}}^{(2)} \overline{\mathbf{d}}^{(1)}+\mathbf{\star}_{\epsilon}^{(1)} \overline{\mathbf{d}}^{(0)}\mathbf{\star}_{\chi^{-1}}^{(3)}\left( \overline{\mathbf{d}}^{(0)} \right)^T \mathbf{\star}_{\epsilon}^{(1)} \right]\\
\times \overline{\mathbf{P}}_A \boldsymbol{A}_{\text{reduced}} = \omega^2\left( \overline{\mathbf{P}}_A \right)^H \mathbf{\star}_{\epsilon}^{(1)} \overline{\mathbf{P}}_A \boldsymbol{A}_{\text{reduced}}, \label{deqn_ex63}
\end{multline}
where the superscript $H$ denotes the Hermitian conjugate. The $\Phi$ eigenvalue equation can be modified similarly, which is
\begin{multline}
\left( \overline{\mathbf{P}}_{\Phi} \right)^H \left( \overline{\mathbf{d}}^{(0)} \right)^T \mathbf{\star}_{\epsilon}^{(1)} \overline{\mathbf{d}}^{(0)} \overline{\mathbf{P}}_{\Phi} \boldsymbol{\Phi}_{\text{reduced}} =\\
 \left( \overline{\mathbf{P}}_{\Phi} \right)^H \omega^2 \mathbf{\star}_{\chi}^{(0)} \overline{\mathbf{P}}_{\Phi} \boldsymbol{\Phi}_{\text{reduced}}. \label{deqn_ex 64}
\end{multline}

\subsection{Perfectly Matched Layer}
The perfectly matched layer (PML) was first proposed by Berenger \cite{Jean1994A} as the absorbing boundary condition in Yee’s algorithm. As the PML became popular, several variations were proposed, and PML was proved to be derivable using the stretched-coordinate approach \cite{1994A}. In \cite{J2014On}, PML has been implemented in the framework of DEC. For conciseness, relative implementation details can be found in Section 4.4 in \cite{J2014On}.

\section{Numerical Examples}

In this section, numerical examples are provided to validate the proposed DEC algorithm for source excitation problems over a broad spectrum. The low-frequency stability and broadband nature of the DEC $\textbf{A}$-$\Phi$ solver is demonstrated as well.

\subsection{Rectangular Wire Loop}

\begin{figure}[!t]
\centering
\subfloat[]{\includegraphics[width=2.5in]{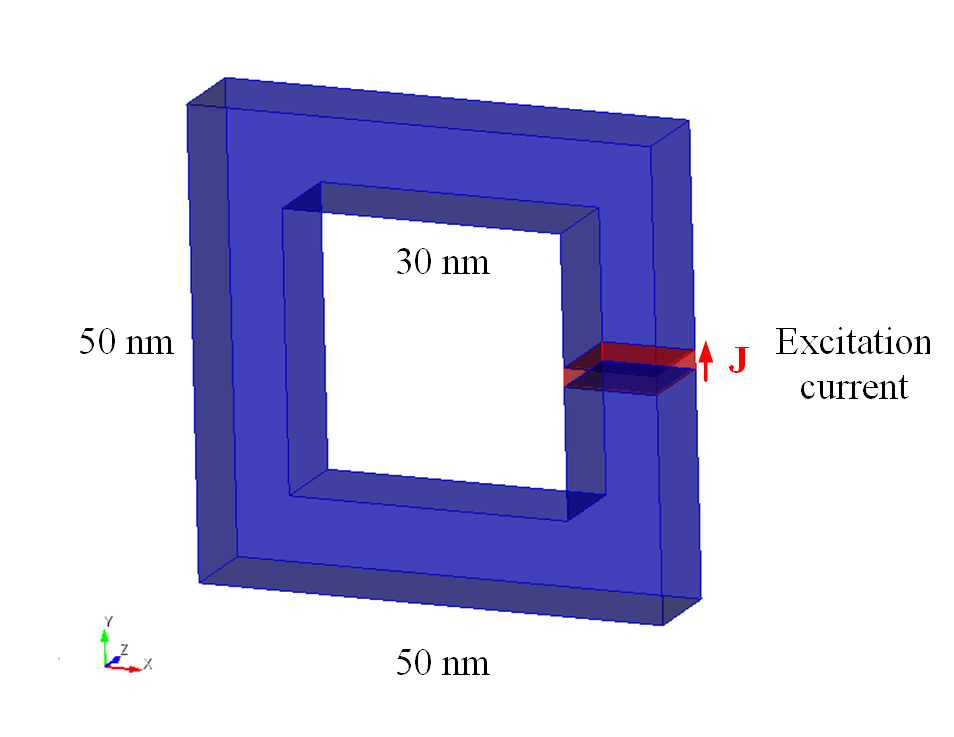}
\label{fig4a}}
\hfil
\subfloat[]{\includegraphics[width=2.5in]{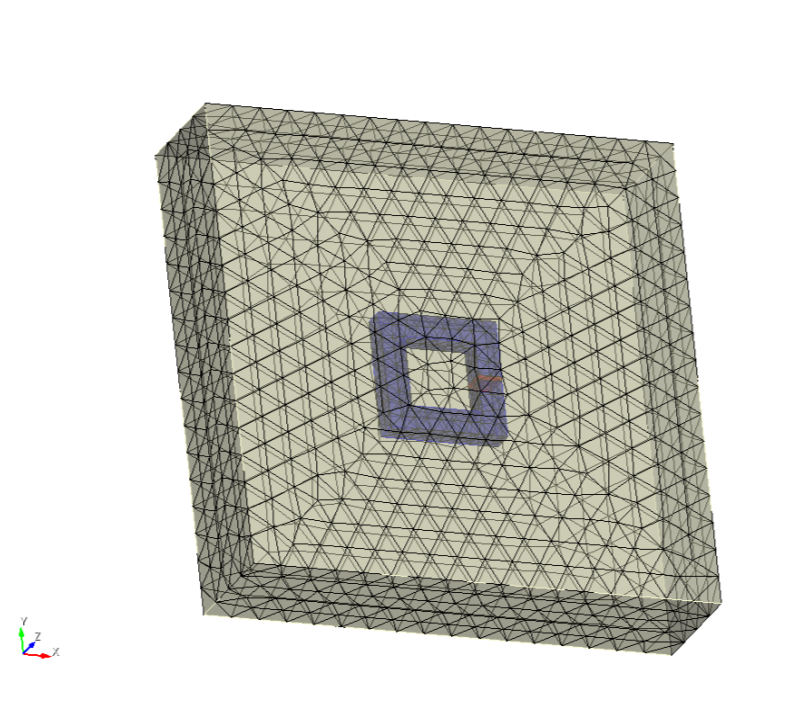}
\label{fig4b}}
\caption{(a) Dimension and (b) tetrahedral mesh of the rectangular wire loop example.}
\label{fig_4}
\end{figure}

\begin{figure*}[!t]
\centering
\subfloat[]{
\includegraphics[width=3.5in]{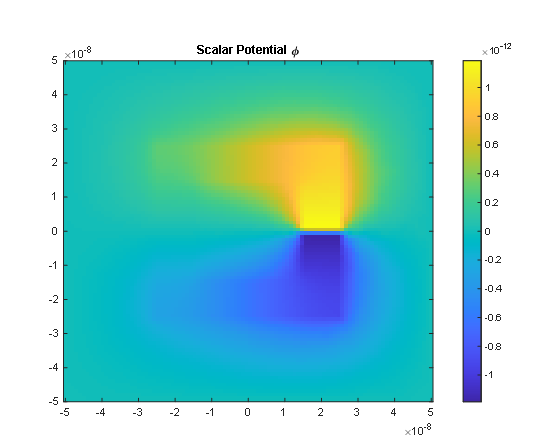}
}
\hfill
\subfloat[]{
\includegraphics[width=3.0in]{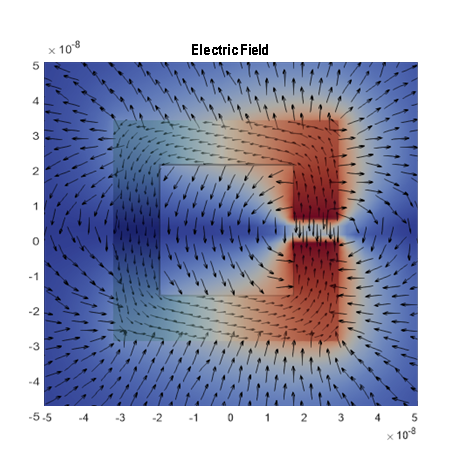}
}
\caption{(a) Scalar potential and (b) electric field results from the proposed DEC solver.}
\label{fig_5}
\end{figure*}

As shown in Fig. \ref{fig_4}(a), where the unit is in nanometers, a rectangular copper wire loop in air is excited by the impressed current $\textbf{J}$ in the excitation gap. The outer size of the wire loop is 50 nm by 50 nm, while the inner size of the loop is 30 nm by 30 nm. The size of the rectangular cross section of the wire is 10 nm by 10 nm. The thickness of the excitation gap is 2 nm. The tetrahedral mesh in the entire calculation domain is shown in Fig. \ref{fig_4}(b). The scalar potential and electric field results obtained using the proposed DEC solver are plotted in Fig. \ref{fig_5}. The scalar potential agrees with expectation, where the impressed current acts as a current source and imposes positive/negative potential on the two ends of the gap. The arrow plot of the electric field reflects the current flow in the conductor as well as the positive/negative charge layers at the gap. After the electric field is extracted, it is trivial to compute the input voltage and current, and thus, calculate the input impedance. Table I summarizes the input impedance, resistance, and inductance of the rectangular wire loop under different frequencies. Also listed are the reference resistance and inductance of the wire loop using approximate formula \cite{2004Inductance}. Clearly, the result obtained from the DEC $\textbf{A}$-$\Phi$ solver is stable from very low frequency to high frequency, which indicates that it is immune to low-frequency breakdown and valid over a broad bandwidth.

\begin{table}[!t]
\begin{threeparttable}
\caption{Calculated Input Impedance $Z$, Resistance $R$, and Inductance $L$ of the Wire Loop\label{tab:table1}}
\centering
\begin{tabular}{|c||c||c||c|}
\hline
Frequency (Hz) & $Z$ ($\Omega$) & $R$ ($\Omega$) & $L$ (H)\\
\hline
1k & $23.88-i2.51 \times 10^{-10}$ & $23.88$ & $4.00 \times 10^{-14}$\\
\hline
1M & $23.88-i2.53 \times 10^{-7}$ & $23.88$ & $4.03 \times 10^{-14}$\\
\hline
100M & $23.88-i2.59 \times 10^{-5}$ & $23.88$ & $4.12 \times 10^{-14}$\\
\hline
1G & $23.88-i2.59 \times 10^{-4}$ & $23.88$ & $4.12 \times 10^{-14}$\\
\hline
10G & $23.88-i2.60 \times 10^{-3}$ & $23.88$ & $4.13 \times 10^{-14}$\\
\hline
\end{tabular}
\end{threeparttable}
\begin{tablenotes}
      \footnotesize
      \item The approximate reference resistance is $ R = 27.84$ $\Omega$ and inductance $L=3.611 \times 10^{-14}$ H.
    \end{tablenotes}
\end{table}

\subsection{Rod Antenna}

\begin{figure}[!t]
\centering
\includegraphics[width=2.5in]{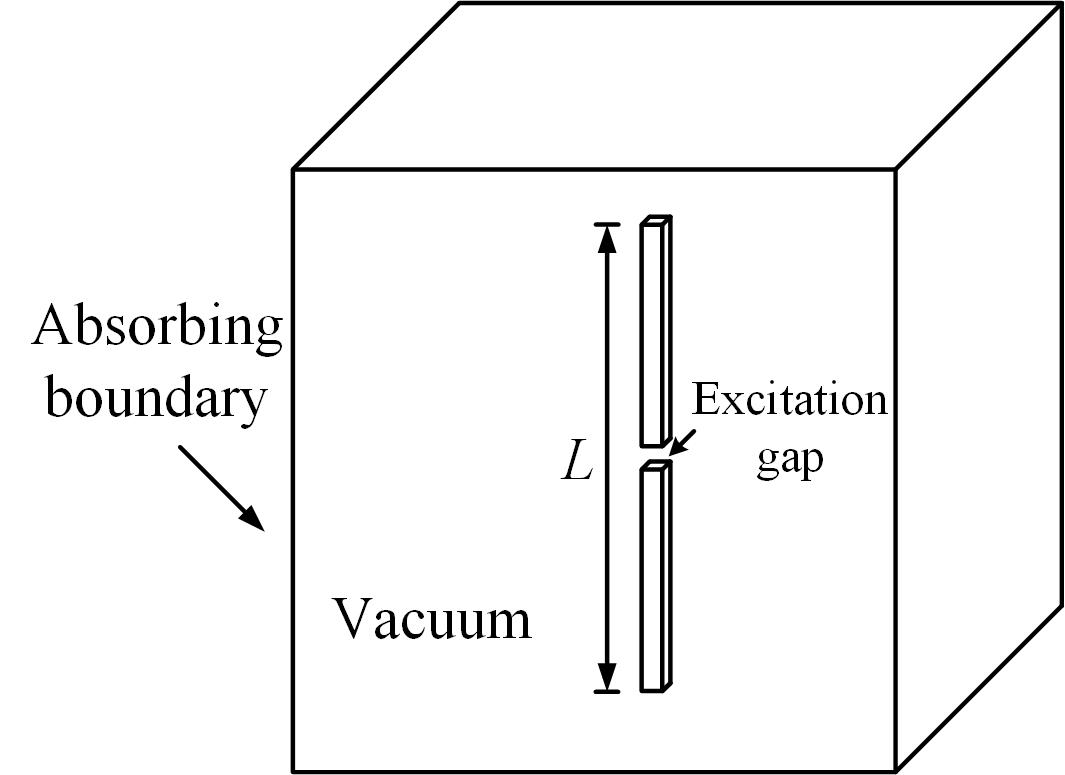}
\caption{An illustration of the rod antenna from \cite{2007FDTD}.}
\label{fig_6}
\end{figure}

\begin{figure}[!t]
\centering
\includegraphics[width=3.5in]{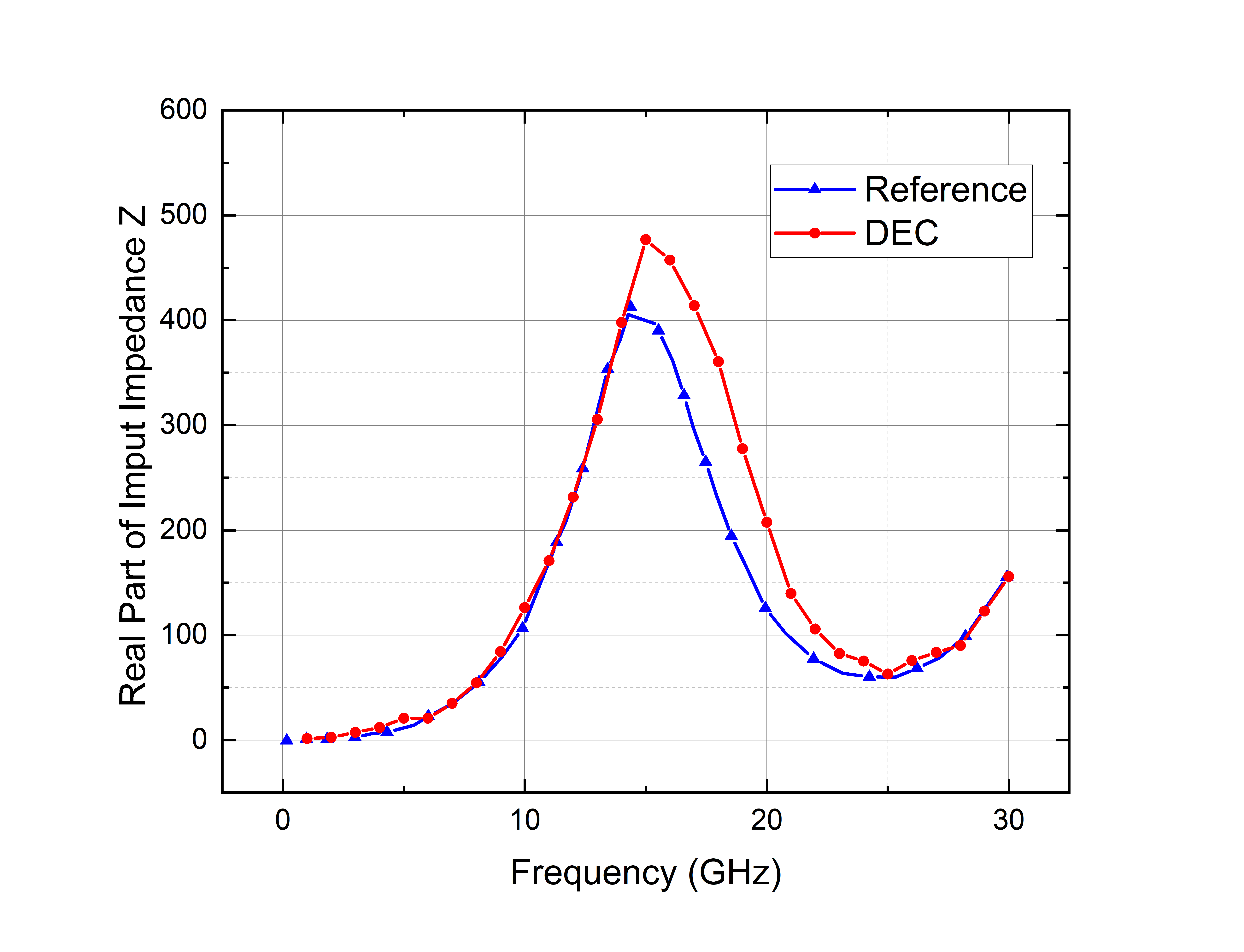}
\caption{Real part of the input impedance $Z$ from \cite{2007FDTD} and DEC.}
\label{fig_7}
\end{figure}

\begin{figure}[!t]
\centering
\includegraphics[width=3.5in]{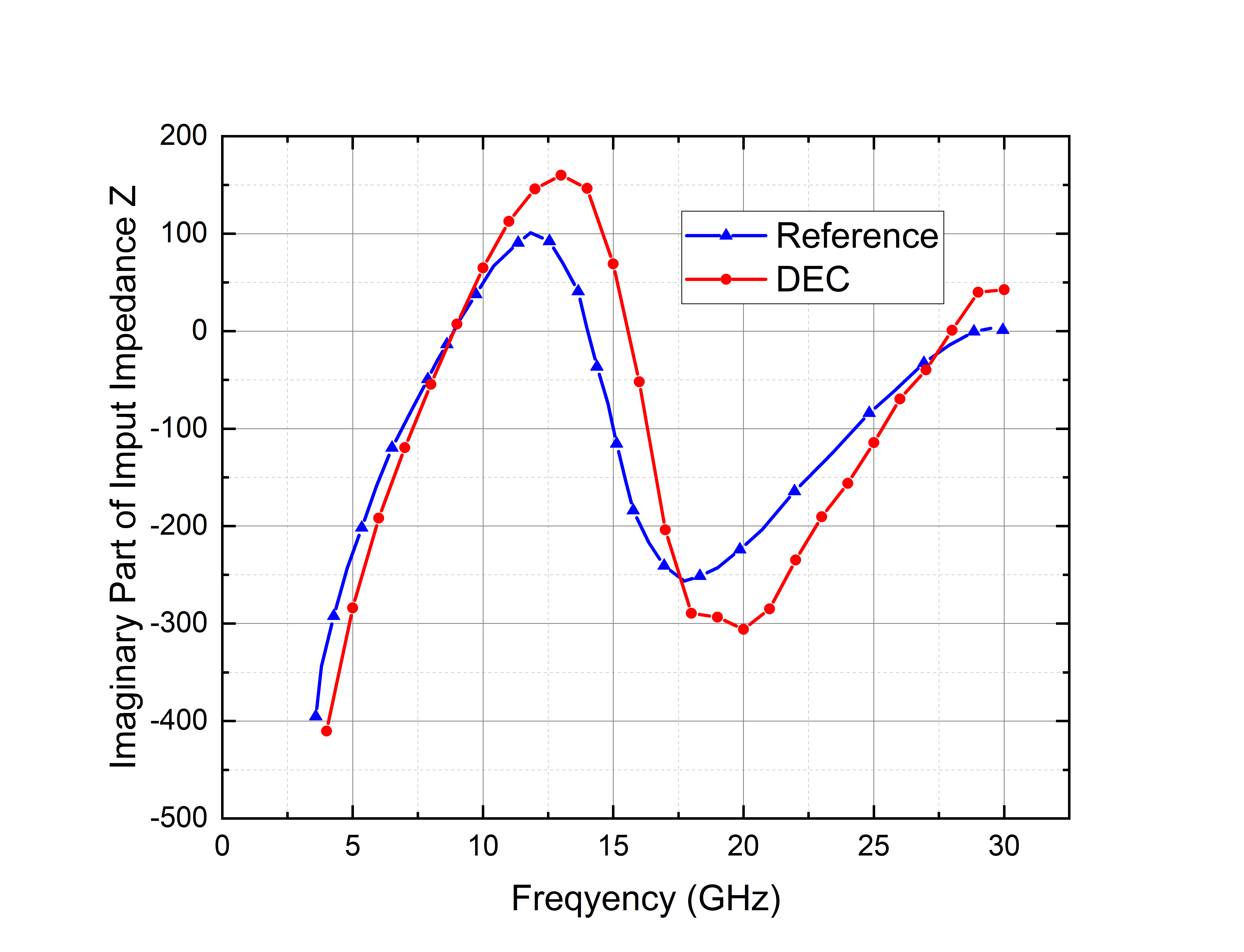}
\caption{Imaginary part of the input impedance $Z$ from \cite{2007FDTD} and DEC.}
\label{fig_8}
\end{figure}

The structure of the rod antenna from \cite{2007FDTD} is shown in Fig. \ref{fig_6}. The rod antenna is composed of two pieces of copper segments with the excitation gap between them. The length of each copper segment is 7.238 mm with cross section 0.517 mm by 0.517 mm. The excitation gap is cube with length 0.517 mm. The total length of the antenna is $L=14.993$ mm. The antenna is placed in cube vacuum region with length 40 mm, and the impedance boundary condition (IBC) is implemented at the boundary as a simple absorbing boundary condition (details on implementing IBC in DEC will be addressed in a separate paper). By impressing excitation current with different frequencies, the input impedance across the excitation gap is calculated over the spectrum. Figs. \ref{fig_7} and \ref{fig_8} show the real part and imaginary part of the input impedance $Z$ obtained from the proposed DEC code and the results from \cite{2007FDTD}. It should be noted that in \cite{2007FDTD}, time domain analysis is performed by using the finite difference time domain method (FDTD) with the perfectly matched layers (PML) as the boundary condtion and a Gaussian impulse as the excitation. With the time domain response, the input impedace in frequency domain is obtained by using the Fourier transform. As can be seen in Figs. \ref{fig_7} and \ref{fig_8}, an overall agreement is observed, despite the minor discrepency due to the abovementioned differences in the two simulations. Nevertheless, the proposed DEC $\mathbf{A}$-$\Phi$ solver can be validated in this rod antenna case with radiating wave physics.

\subsection{Broadband Stability}

\begin{figure}[!t]
\centering
\includegraphics[width=3.5in]{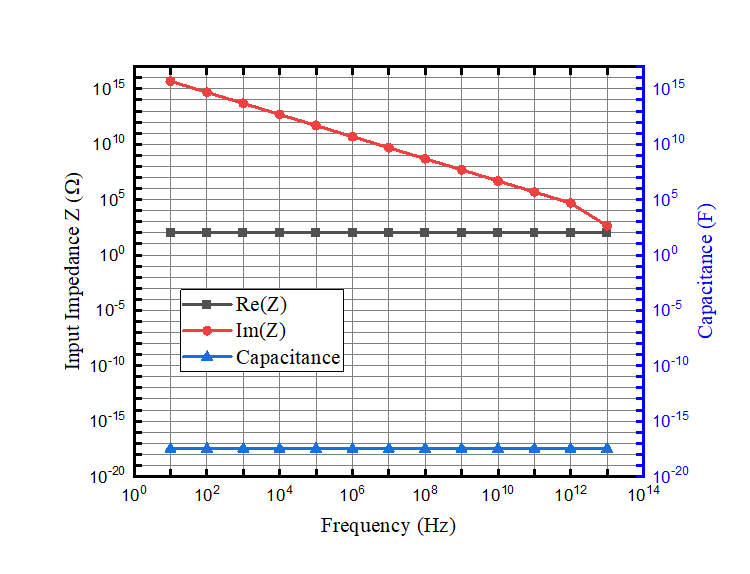}
\caption{The extracted input impedance, resistance and capacitance of the nano-scale rod antenna from 10 Hz to 10 THz.}
\label{fig_9}
\end{figure}

In order to further explore the broadband stability of the proposed DEC $\mathbf{A}$-$\Phi$ solver, the size of the rod antenna in previouse case is reduced to nano-scale. In this case, the length of the rod antenna is $L=500$ nm with cross section 10 nm by 10 nm. The length of the excitation gap is 10 nm. The vacuum cube box is of  length 1000 nm with PEC boundary condition for simplicity. Fig. \ref{fig_9} shows the extracted input impedance, resistance and capacitance of the nano-scale rod antenna within a very broad spectrum, from 10 Hz to 10 THz. It can be observed clearly that the extracted results are very stable over the extremely wide spectrum. Note that when the frequency is around 10 THz, it is close to the resonance frequency of the rod antenna, and the imaginary part of the input impedance starts to oscillate. To our best knowledge, there is no other work that simulates such a broadband problem and thus there is no reference result to compare. This numerical example is intended to demonstrate the broadband stability of the proposed solver from DC to infrared frequency range.

\section{Conclusion}

In this paper, a DEC $\textbf{A}$-$\Phi$ solver is proposed for broadband and multi-scale analysis in computational electromagnetics. The derivation of the $\textbf{A}$-$\Phi$ formulation along with the generalized Lorenz gauge are provided, and the detailed implementation procedure of DEC in the $\textbf{A}$-$\Phi$ formulation is addressed as well. Numerical examples for a rectangular wire loop and a rod antenna with impressed current excitation problem using the proposed DEC $\textbf{A}$-$\Phi$ solver are presented as validations of the proposed algorithm for both quasi-static and wave physics scenarios, as well as its broadband stability. More importantly, the fact that the $\textbf{A}$-$\Phi$ formulation is immune to low-frequency breakdown is demonstrated in the example. Stokes’ theorem, Gauss's theorem and charge conservation are naturally preserved in the proposed DEC solver, and unstructured mesh schemes such as tetrahedral meshes can be utilized in the proposed solver easily. Overall, the proposed DEC $\textbf{A}$-$\Phi$ solver can serve as an efficient broadband and multi-scale solver in computational electromagnetics.

\section*{Acknowledgments}
This work is supported by the Consortium for Electromagnetic Science and Technology at Purdue University. The authors wish to thank the member companies: Synopsys, Cadence, Siemens and Ansys, for their support.

%

%
%
%
%
%
%
%
%
%

\end{document}